\documentclass[a4paper, twocolumn]{article}
\usepackage[a4paper, margin=1in]{geometry}
\pdfoutput=1
\usepackage[utf8]{inputenc}
\usepackage{xcolor}
\usepackage{textcomp}
\usepackage[numbers]{natbib}
\usepackage{hyperref}
\usepackage{titlesec}
\usepackage{fancyvrb}
\usepackage[parfill]{parskip}
\usepackage[affil-it]{authblk}

\titleformat*{\section}{\large\bfseries}

\renewenvironment{abstract}{%
  \centerline{\large\bf Abstract}
    \begin{quote}}
    {\par\end{quote}\vspace{.5em}}

\title{\bf Decomposing reverse-mode automatic differentiation}

\author{Roy Frostig}
\author{Matthew J. Johnson}
\author{Dougal Maclaurin}
\author{Adam Paszke}
\author{Alexey Radul\thanks{Google Research \\\\ \textit{Presented at LAFI 2021 at POPL, 17 January 2021}}}
\affil{}
\date{}

\newcommand{\vocab}{\emph}
\newcommand{\ttt}{\texttt}

\def\preverba{}
\def\postverba{}

\begin{document}

\maketitle

\begin{abstract}
  We decompose reverse-mode automatic differentiation into
  (forward-mode) linearization followed by transposition.
  Doing so isolates the essential difference between
  forward- and reverse-mode AD, and simplifies their joint
  implementation. In particular, once forward-mode AD rules are defined
  for every primitive operation in a source language,
  only \emph{linear} primitives require an additional transposition rule
  in order to arrive at a complete reverse-mode AD implementation.
  This is how reverse-mode AD is written in JAX
  \citep{jax2018github, frostig2018compiling}
  and Dex
  \citep{maclaurin2019dex, paszke2021pointful}.
\end{abstract}

Automatic differentiation (abbreviated AD or autodiff) comes in
several ``modes''---algorithmic variants that present an
efficiency tradeoff.
The most common two are \vocab{forward-} and \vocab{reverse-mode} AD.
At their core, these compute slightly different functions:
forward-mode AD carries out a Jacobian-vector product (JVP),
while reverse-mode AD carries out a vector-Jacobian product (VJP),
both for some vector taken as input.
To produce a Jacobian matrix, one might invoke either
mode over each standard basis vector.

Naturally, the two modes share apparent similarities,
and maintainers of AD tools might wonder about
any redundancy to avoid or to exploit in their implementation.
Typically, implementing either mode requires defining a corresponding
\vocab{transfer rule} for each base primitive operation in the source language
(e.g.,\ for element-wise array division, for matrix multiplication,
for array slicing, etc). Does maintaining both modes require two
complete, distinct collections of transfer rules?
The thought that it might has led industry-grade tools such as
PyTorch \citep{paszke2019pytorch} to offer only one mode (reverse).
JAX \citep{jax2018github, frostig2018compiling}
and Dex \citep{maclaurin2019dex, paszke2021pointful}
offer both modes by reusing the forward-mode transfer rules in reverse-mode AD
according to this note.

Our contribution is to decompose reverse-mode AD into two steps:
\vocab{linearization} and \vocab{transposition}.
Linearization constructs a JVP function, by standard
forward-mode AD.  It will be important that this JVP consist of only \vocab{linear}
primitives, which we ensure with \vocab{partial evaluation},
a standard general-purpose compiler technique.
Transposition corresponds to the mathematical notion of
transposition of a linear map, which we implement as a new program transformation
that boils down to transposing linear primitives.
Transposition is an interesting operation in its own right---the
class of linear programs can be viewed as a flexible representation for structured sparsity of linear maps,
and this program transformation achieves mathematical transposition
of the matrix without destroying the sparsity pattern.

This decomposition underscores %
that transposition is what fundamentally relates forward-
and reverse-mode, as the terms JVP and VJP suggest.
A monolithic VJP transformation must:
(1) linearize primitive operations,
(2) compute the primal output while storing intermediate values, and
(3) reverse the order of operations.
We instead treat each of these steps as a standalone transformation.
The linearization step ensures that transposition rules
need only be defined for linear primitives in order to arrive at
a complete AD system.

\section{Reverse from forward}

For clarity, we will use a Haskell-like notation for types in the sequel.
We use the type constructor \ttt{T} for ``tangent space'', writing \ttt{T a} for the tangent space of \ttt{a}.  We do not distinguish tangent spaces from cotangent spaces, because the ones that
can occur in a realizable computation are canonically isomorphic.

In this notation, the %
\ttt{jvp} and \ttt{vjp} transformations have the types:

\preverba{}
\begin{Verbatim}[samepage=true]
  jvp :: (a -> b) -> (a, T a) -> (b, T b)
  vjp :: (a -> b) -> (a, T b) -> (b, T a)
\end{Verbatim}
\postverba{}

In both cases, the returned functions still compute the primal mapping \ttt{f :: a -> b} (first component of the argument and result), but they are additionally augmented with a linear transform over the tangent types (given by the second input and result).
While \ttt{jvp} computes the directional derivative of the original mapping, \ttt{vjp} computes the transpose, namely directional gradients (as the name suggests).
We capture the similarity between \ttt{jvp} and \ttt{vjp} with a new function transformation we call
\ttt{transpose}, with type:

\preverba{}
\begin{Verbatim}[samepage=true]
  transpose :: (a -o b) -> (b -o a)
\end{Verbatim}
\postverba{}

Here the \ttt{-o} means a \emph{structurally linear} function from \ttt{a} to \ttt{b}.  We do not give a precise definition of ``structurally linear'' in this note, but the idea is that the function should be implemented with only linear primitives, and should preserve linearity through all sub-computations by using the input exactly once in each additive term of the output.

The operation of \ttt{transpose} is to accept a structurally linear function that implements a linear map, and return a structurally linear function representing the transpose of the same map.
It has been shown that this can be derived automatically, using techniques similar to AD \citep{piponi2009two}.
To take advantage of this, we need to isolate the transposable linear map from \ttt{jvp},
by slightly reformulating as \ttt{linearize}:

\preverba{}
\begin{Verbatim}[samepage=true]
  linearize ::
      (a -> b) -> a -> (b, T a -o T b)
\end{Verbatim}
\postverba{}

The difference from \ttt{jvp} is that \ttt{linearize} does not compute the primal and derivative functions at the same time, but instead only computes the primal part and returns the linear derivative function as a result.
Fortunately, the difference in type signatures between \ttt{jvp} and \ttt{linearize} is just the effect of \vocab{partially evaluating} \ttt{jvp f} with respect to the \ttt{a} argument, relying on the fact that the primal result (of type \ttt{b}) depends only on the primal input and not on the direction of differentiation.

Partial evaluation is a well-studied technique used in optimizing compilers.
In brief, given the code of some function \ttt{g}, and a known value for one of its inputs, we can, at function transformation time, compute all the intermediate values that depend on just the known input (in this case, the primal point).
The remainder of the computation we stage into a new, generally simpler and faster---and, in this case, linear---function that accepts the remaining arguments of \ttt{g}.

That last comment about linearity is important,
and is a critical thing we gain from true partial evaluation as opposed to just closing over the \ttt{a} input.
The partial evaluation computes and stores everything that depends only on the primal, and the nature of differentiation implies that all the residuals are, in fact, structurally linear in exactly the way they need to be for \ttt{transpose} to work.
Partial evaluation is of course also critical for the performance of reverse-mode AD, because it prevents potentially wasteful recomputation of the primal values on the reverse pass.
To summarize:

\preverba{}
\begin{Verbatim}[samepage=true]
  linearize f =
      let ft = jvp f
       in partial_eval ft
\end{Verbatim}
\postverba{}

With \ttt{linearize} in hand we can define \ttt{vjp} concisely:
\preverba{}
\begin{Verbatim}[samepage=true]
  vjp f (x, ty) = 
      let (y, ft) = linearize f x
          ftt     = transpose ft
       in (y, ftt ty)
\end{Verbatim}
\postverba{}
and we have achieved our goal of writing differentiation rules for our primitives only once.
We still need to provide a \ttt{jvp} rule for each basic operation in our system, but the rule set necessary to implement \ttt{transpose} is significantly smaller, as it only considers the linear operations.
The \ttt{partial\_eval} transformation can be implemented generically for all operations---the usual interpretation is that the output is known if and only if all inputs are known.
The only place where care needs to be taken is around control flow constructs, but there are usually very few in any language.
In fact, many tracing mechanisms used for AD erase control flow entirely. %

\section{Related work and discussion}

The decomposition laid out in this note may be known in, or implied by,
folklore among some AD experts, but we have not found it in writing,
and neither have many of our colleagues in the research community.
This has been the implementation approach in JAX since before the project's
initial open-source release in 2018.
Several AD tools implement only one mode
\citep{abadi2016tensorflow, paszke2019pytorch},
or implement both modes using two rule sets
\citep{baydin2016diffsharp, innes2019zygote}.
Work by Elliott touches peripherally on transposition in
\citep{elliott2018essence} in its discussion of duality and gradients.
AD surveys and textbooks \citep{baydin2017survey, griewank2008evaluating}
commonly observe the fundamental bit that forward-mode AD
computes a JVP and that reverse-mode AD computes its transpose (a VJP).

It is tempting to try to quantify the cost savings due to
a smaller rule set.
In JAX the number of transposition rules is roughly
40\% of the JVP rule count.
Still, such a quantity should be read qualitatively:
it depends on the choice of language primitives, whose design is
determined by concerns extrinsic to AD.
For instance, JAX offers just-in-time compilation, affording it a minimal
primitive set of roughly 120 operations (compare to $\sim$400 primitives
in PyTorch, and several hundreds or more in TensorFlow \citep{abadi2016tensorflow}).
Additionally, classes of similar primitives' rules are merely instances of
a common rule template (e.g.\ for binary operations that are linear in both operands);
here we have counted them individually.
Some primitives (e.g.,\ division) are only linear with respect to some of their operands, obscuring the notion of a precise rule count.
Finally, fewer rules are not always simpler rules.

\clearpage

\bibliographystyle{alpha}
\bibliography{ref}

\end{document}